\documentstyle[12pt,epsf]{article}
\newcommand{\be}{\begin{equation}} 
\newcommand{\en}{\end{equation}}
\newcommand{\bea}{\begin{eqnarray}}
\newcommand{\ena}{\end{eqnarray}}

\newcommand{\hbo}{\hbox to 1 true cm {\hfill } } 
\newcommand{\tr}{\hbox{tr}}

\input psfig 
\def\dslash{\partial\kern-.5em\slash}
\def\kslash{k\kern-.5em\slash}
\def\pslash{p\kern-.5em\slash}

\begin{document} 
\vglue 1truecm
  
\vbox{ UNITU-THEP-8/97
\hfill April 28, 1997
}
  
\vfil 
\centerline{\bf Improving the signal--to--noise--ratio in lattice gauge 
theories } 
\bigskip
\centerline{ Kurt Langfeld, Hugo Reinhardt and Oliver Tennert } 
\bigskip
\centerline{ Institut f\"ur Theoretische Physik, Universit\"at 
   T\"ubingen }
\centerline{D--72076 T\"ubingen, Germany.}
\vskip 1.5cm
  
\begin{abstract}
Renormalization of composite fields is employed to suppress 
the statistical noise in lattice gauge calculations. We propose a 
new action which differs from the standard Wilson action by ''irrelevant'' 
operators, but suppresses the fluctuations of the plaquette. 
We numerically study the Creutz ratios and find a scaling window. 
The SU(2) mass gap is estimated. We prove that the contributions of 
the ''irrelevant'' operators to the screening mass decrease 
towards the continuum limit. The results obtained from the action with 
noise suppression are compared with those of the standard Wilson action. 

\end{abstract} 

\vspace{1cm} 
\begin{center} 
{\it renormalization of composite fields, lattice gauge theory, \\ 
signal--to--noise--ratio }
\end{center} 

\vfil
\hrule width 5truecm
\vskip .2truecm
\begin{quote} 
$^*$ Supported in part by DFG under contract Re 856/1--3. 

\end{quote}
\eject
\section{Introduction} 

The present lattice calculations provide the only rigorous 
approach to low energy Yang-Mills theories. After the scaling window 
had been discovered by Creutz in his pioneering work~\cite{creu80}, 
lattice simulations provided first informations on the ratio 
of the low lying glue-ball masses and the string tension for the 
SU(2)~\cite{tep87} and the SU(3)~\cite{berg86,sesam} gauge group. 

Unfortunately, the limited capacity of the computers put severe 
constraints on the accuracy of ''lattice measurements''. Firstly, 
the finite number of lattice links correspond to a finite physical 
volume. Nowadays, a physical volume of $(1.6 \, \hbox{fm})^4$ 
is available for reasonable values of the lattice spacing (see 
e.g.~\cite{sesam}). Secondly, the finite number of independent 
configurations which are employed to calculate the expectation value 
of the desired operator implies that the ''lattice measurement'' 
is contaminated with statistical noise. 

Weisz and Symanzik have shown that the situation corresponding to the 
finite size problem can be significantly improved by using an 
improved action~\cite{sym83}. In the numerical simulation, 
an effective action is used which already contains corrections 
from perturbative radiation. In recent years, much work has been 
devoted to the development of such improved lattice actions, which are 
often referred to as ''perfect'' lattice actions~\cite{has94}. 

In this paper, we will focus on the noise problem. In order to outline 
the conceptual nature of the noise problem, we briefly review the 
arguments presented in~\cite{sch90}. Glue-ball (screening) masses $m_g$ 
are extracted from correlation functions, i.e. 
\be 
C(t):= \langle \Phi (t) \, \Phi (0) \rangle \; \approx \; const. \; 
e^{-m_g t} \; , 
\label{eq:1} 
\en 
where the brackets indicate an average over independent lattice 
configurations. The statistical error of the desired quantity 
is measured by the standard deviation~\cite{sch90} 
\be 
\langle \Phi (t) \Phi (0) \Phi (t) \Phi (0) \rangle \; - \; 
C^2(t) \; \approx \; \langle \Phi ^2 (0) \rangle \; . 
\label{eq:2} 
\en 
From perturbation theory, one knows that composite operators 
acquire new divergences implying that the statistical error 
of the correlation function $C(t)$, i.e. $\sqrt{\langle \Phi ^2 (0) 
\rangle}$, diverges in the continuum limit $a \rightarrow 0$. 
The disastrous and fundamental problem therefore is that the 
signal--to--noise--ratio vanishes in the continuum limit. 

In the recent past, two concepts have been established to be important in 
order to improve the signal--to--noise--ratio. Firstly, the choice of a 
non-local operator $\Phi(x)$ in (\ref{eq:1}) (which nevertheless 
carries the quantum numbers of the state under investigation) might 
result in a composite operator $\Phi ^2(x)$ which is free of 
ultra-violet divergences ({\it smearing}~\cite{ape87}). Secondly, 
informations from the links of a former update step is used 
to enhance the signal ({\it fuzzing}~\cite{tep86}). Hybrid algorithms, 
which combine ''smearing'' and ''fuzzing'', as well as an estimate 
of their impact on the signal--to--noise--ratio can be found 
in~\cite{sch90}. 

In this paper, we prose a new method to improve the 
signal--to--noise--ratio. 
The method is inspired from the renormalization procedure 
for composite operators in continuum quantum field theory. 
The basic idea is to add to the Wilson action an additional term 
which vanishes faster than the Wilson action in the continuum limit, i.e. 
we add an ''irrelevant operator'', but which suppresses the statistical 
noise. We study the efficiency of our method by calculating the 
SU(2) mass gap employing the ''old'' idea of plaquette--plaquette 
correlations. 

The paper is organized as follows: in the next section, we briefly 
review the renormalization of composite operators in continuum 
quantum field theory. We then discuss the modifications of the 
Wilson action by ''irrelevant'' composite fields which 
yield the suppression of the noise. In the third section, we present 
our numerical results. Conclusions are left to the final section.

\section{ Renormalization of composite operators } 

\subsection{ In continuum quantum field theory } 

For illustration purposes, we here consider the continuum 
quantum field theory of a field $\phi (x)$ which is described 
by the Euclidean partition function 
\be 
Z[\eta ](g) \; = \; \int {\cal D} \phi (x) \; \exp 
\left\{ -S[\phi ](g) + \int d^4x \, \eta (x) \phi (x) \right\} \; , 
\label{eq:3} 
\en 
where a regularization is understood in order to make (\ref{eq:3}) 
well defined. 
For simplicity, we assume that the Euclidean action $S$ contains only 
one parameter $g$ (e.g. the coupling strength). The external source $\eta $ 
linearly couples to the field $\phi (x)$ implying that functional 
derivatives of $Z[\eta ](g)$ with respect to $\eta $ yield connected 
Green's functions 
\be 
\langle T \, \phi (x_1) \ldots \phi (x_N) \rangle \; .
\label{eq:4} 
\en 
These Green's functions are generically divergent in four space-time 
dimensions, if the regulator is removed. Renormalized Green's functions 
are obtained from the generating functional 
\be 
Z_R[\eta _R ](g_R) \; := \; Z[ Z_\eta \eta _R ](Z_g g_R)  
\label{eq:5} 
\en 
by performing the functional derivative with respect to $\eta _R$. 
The $Zs$ are the so-called renormalization constants, and $\eta _R$ 
is the renormalized source which accounts for field renormalization, 
and $g_R$ is the renormalized parameter. In order to guarantee that 
(\ref{eq:5}) yields finite Green's functions, we have tacitly assumed that 
the field theory (\ref{eq:3}) is multiplicative 
renormalizable~\cite{ch84}, i.e. all divergences can be absorbed 
in the renormalization constants $Z$. 

The crucial observation is that, if we allow for composite field 
insertions, i.e. if we are interested in the limit $x_1 \rightarrow 
x_2$ in the renormalized Green's function, new divergences 
arise in $Z_R[\eta _R ](g_R)$. In order to renormalize these insertions, 
we generalize 
\bea 
&& Z_R[\eta _R ](g_R) \rightarrow Z_R[\eta _R, j_R](g_R) \; = \; 
\label{eq:6} \\ 
&&
\int {\cal D} \phi (x) \; \exp \left\{ -S[\phi ](Z_r g) + \int d^4x \, 
\left[ Z_\eta \eta_R (x) \phi (x) + Z_j j_R(x) \phi ^2(x) \right] 
\right\} \; . 
\nonumber 
\ena 
The additional divergences due to the composite field $\phi ^2(x)$ 
can be absorbed in the renormalization constant $Z_j$. 

The dependence of the renormalization constants on the regulator 
is of course not known a priori. Perturbation theory usually 
provides a systematic way to extract this dependence~\cite{ch84}. 
In the context of numerical lattice gauge calculations, 
the question arises, how one should choose the bare source $j(x)$ 
as function of the lattice spacing in order to 
renormalize the composite field insertions and 
therefore to reduce the statistical noise in the continuum limit. 
We will answer this question in the next section.

\subsection{ In lattice gauge calculations } 

The partition function of SU(2) lattice Yang-Mills theory is defined 
as a functional integral over the link variables $U_\mu (x)$, i.e. 
\be 
Z \; = \; \int {\cal D} U_\mu \; \exp \{ -S \} \; , 
\label{eq:7} 
\en 
where the standard Wilson action is given by 
\be 
S_W \; = \; \sum _{ \{x\} \mu \nu } \beta \, \left[ 1 \, - \, 
P_{\mu \nu }(x) \right] \; . 
\label{eq:8} 
\en 
$P_{\mu \nu }$ is the plaquette, which is built from four 
link variables, i.e. 
\be 
P_{\mu \nu }(x) \; = \; \frac{1}{2} \tr \left\{ 
U_{\mu }(x) U_{\nu }(x +\mu ) U^\dagger _\mu (x +\nu ) 
U^\dagger _\nu (x) \right\} \; . 
\label{eq:8a} 
\en 
The functional integral (\ref{eq:7}) is defined 
on a lattice with lattice spacing $a$, which serves as the 
ultraviolet regulator. In the continuum limit $(a \rightarrow 0)$, 
the plaquette is 
\be 
P_{\mu \nu }(x) \; = \; 1 \, - \, \frac{a^4}{4} F^{a}_{\mu \nu } 
F^{a}_{\mu \nu } \; + \; {\cal O}(a^6) \; , 
\label{eq:9} 
\en 
where $F^a_{\mu \nu }$ is the usual field strength tensor. 
In this paper, we will confine ourselves to the plaquette-plaquette 
correlation function $\langle P_{\mu \nu }(x) P_{\alpha \beta }(0) 
\rangle $ in order to extract the mass gap of the SU(2) 
lattice theory. From (\ref{eq:2}), it is clear that this 
correlation function is plagued by a statistical noise which 
diverges in the continuum limit. From the discussions in the last 
section, it is now evident that one must add a term 
$\sum _{ \{x\} \mu \nu } j(x) P_{\mu \nu }^2(x)$ with a suitable choice of 
$j(x)$ to the action (\ref{eq:8}) in order to avoid this divergence. 
We here propose to perform the numerical simulation using the action 
\be 
S \; = \; \sum _{ \{x\} \mu \nu } \beta \, \left[ 1 \, - \, 
P_{\mu \nu }(x) \right]  
\; + \;  \sum _{ \{x\} \mu \nu } j \, \left[ P_{\mu \nu }(x) - 
{\cal A} \right] ^2 \; , 
\label{eq:10} 
\en 
where $j$ and ${\cal A}$ are constants. In fact, we will choose ${\cal A}$ 
to be the average value of the plaquette $P_{\mu \nu }$, which will be a 
function of $\beta $ and $j$. 

Let us study the naive continuum limit of the action $S$ (\ref{eq:10}). 
Using (\ref{eq:9}), a direct calculation yields (up to a constant) 
\be 
S \; = \; \left[ \beta \, - \, 2j \, (1 - {\cal A}) \right] \; 
\frac{a^4}{4} F^a_{\mu \nu } F^{a}_{\mu \nu } \; + \; 
{\cal O}(a^6) \; . 
\label{eq:11} 
\en 
This implies that the action $S$ cannot be distinguished in the naive 
continuum limit from Wilson's action (\ref{eq:8}) with an 
effective parameter $\beta _{eff} = \beta - 2j (1- {\cal A}) $. 
In the quantum continuum limit $(\beta \rightarrow \infty )$, the 
average plaquette and effective inverse temperature $\beta _{eff}$ are 
approximately given by~\cite{creu80} 
\be 
{\cal A} \; = \; 1 \, - \, \frac{3}{4 \beta } \; , \hbo 
\beta _{eff} \; = \; \beta \, - \, \frac{3j}{2 \beta } \; . 
\label{eq:12} 
\en 
If $j$ increases less than linearly with $\beta $, the results of the 
quantum theory using $S$ should agree with those which are obtained 
by employing the Wilson action. 

On the other hand, the term in (\ref{eq:10}) proportional to $j$ 
further constrains the plaquette to its average value ${\cal A}$ 
and therefore suppresses statistical fluctuations around the 
average value of the plaquette. 
The key point is that the action $S$ (\ref{eq:10}) differs from the 
Wilson action by ''irrelevant'' operators which are chosen to suppress 
the statistical noise of the plaquette. 
The prize one has to pay is that the average plaquette value 
must already be known at the beginning of the numerical simulation.

\section{Numerical results } 

\subsection{ Creutz ratios } 

\begin{figure}[t]
\centerline{ 
\epsfxsize=8cm
\epsffile{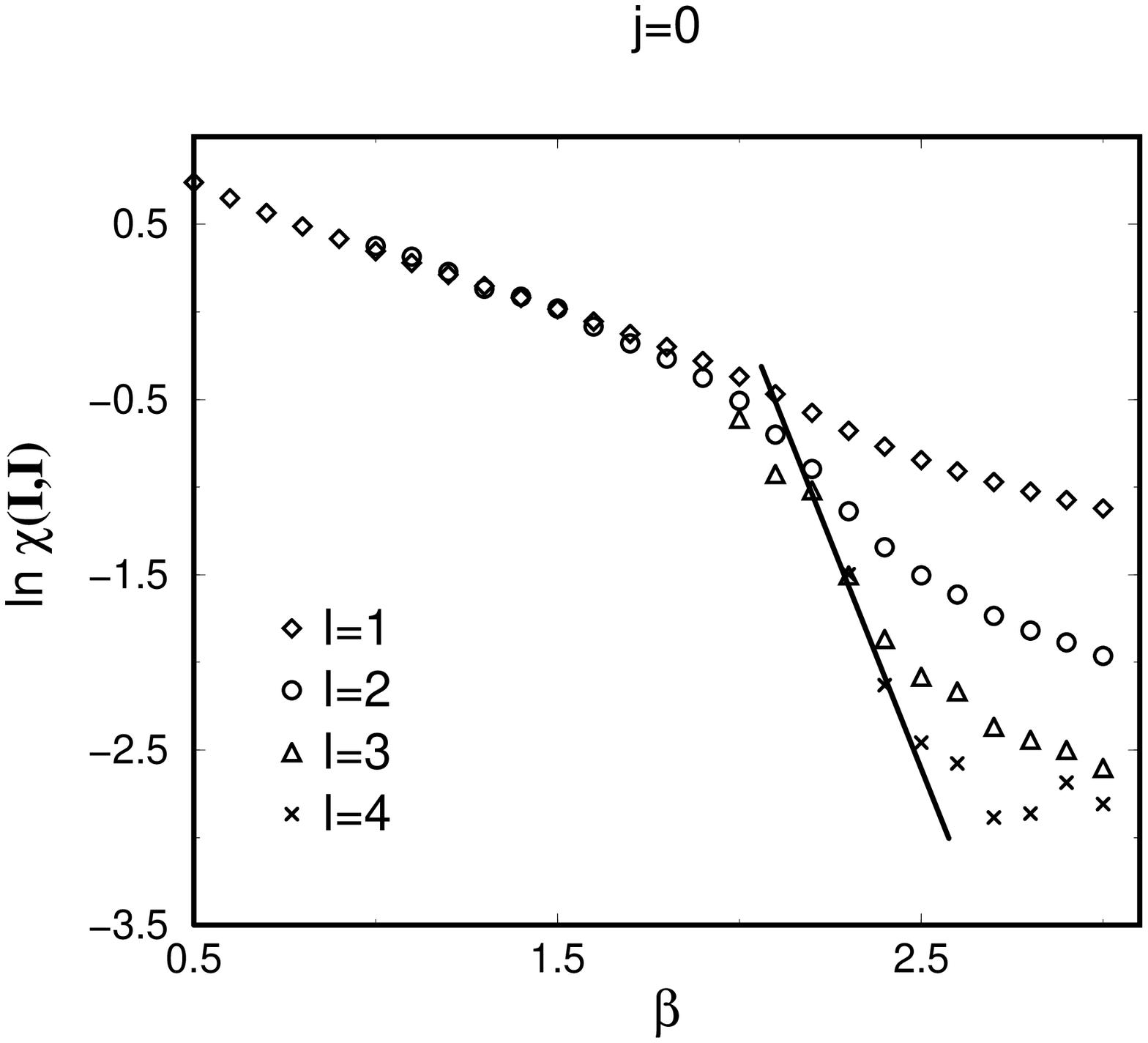} 
\epsfxsize=8cm
\epsffile{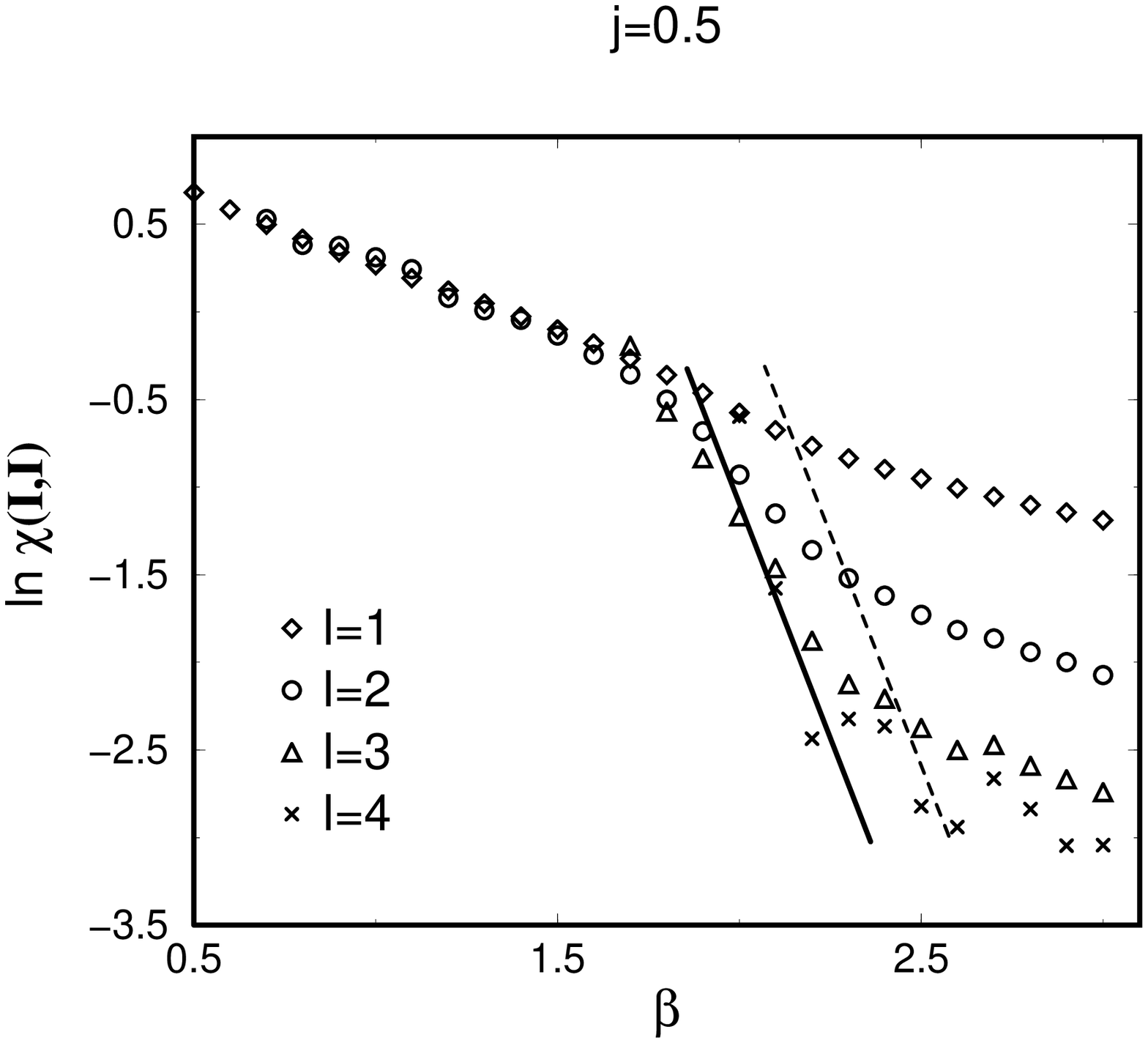} 
}
\vspace{-.8cm} 
\caption{ The Creutz ratios employing the standard Wilson action 
   (left) and using the action $S$ (\protect{\ref{eq:10}}) with $j=0.5$ 
   (right). The dashed line in the right hand picture indicates 
   the perturbative scaling behavior of the case $j=0$. }
\label{fig:1} 
\end{figure} 
We perform our numerical simulations of the quantum theory 
employing the action (\ref{eq:10}) on a $10^4$ lattice using the heat 
bath algorithm by Creutz~\cite{creu80}. Our purpose is to  
demonstrate the mechanism of noise suppression proposed in the previous 
sections, rather than to provide new precision 
measurements. In the latter case, one should resort to ''improved'' 
actions~\cite{sym83} as well as a larger number of lattice points. 

The first task is to calculate the average plaquette ${\cal A}$ 
self-consistently for given values for $\beta $ and $j$. 
We apply the following procedure (we leave it to the reader 
to develop his own method): before the lattice has reached 
its thermo-dynamical equilibrium, we use the lattice average 
from the previous heat bath step for ${\cal A}$. When equilibrium is 
reached, we calculate the average plaquette taking into account 
all heat bath steps. In a particular step, we assign the actual value 
of this average plaquette to ${\cal A}$. We then perform a large 
number of heat bath steps to obtain an accurate value of ${\cal A}$, 
which subsequently enters the numerical calculations of correlation 
functions, where a smaller number of heat bath steps is sufficient. 

The crucial question is whether the scaling limit is reached for finite 
values of $j$. In order to answer this question, we calculate the 
Creutz ratios~\cite{creu80} as a function of $\beta $. The left 
picture of figure \ref{fig:1} shows the case $j=0$. These are the 
Creutz ratios which one obtains using the standard Wilson 
action. These results are compared with those for $j=0.5$ (in the 
right picture of figure \ref{fig:1}). The lines indicate 
the scaling behavior which has been calculated with the help of 
the perturbative renormalization group $\beta $-function. 
The crucial observation is that the model with non-vanishing source 
$j$ also approaches the scaling behavior (shown by the lines in figure 
\ref{fig:1}) which is predicted by the perturbative renormalization 
group. Also note that the perturbative scaling already sets in at smaller 
values of $\beta $ compared with those of the case $j=0$. This 
is precisely what one expects from (\ref{eq:12}).

\subsection{ Noise suppression } 

\begin{figure}[t]
\centerline{ 
\epsfxsize=12cm
\epsffile{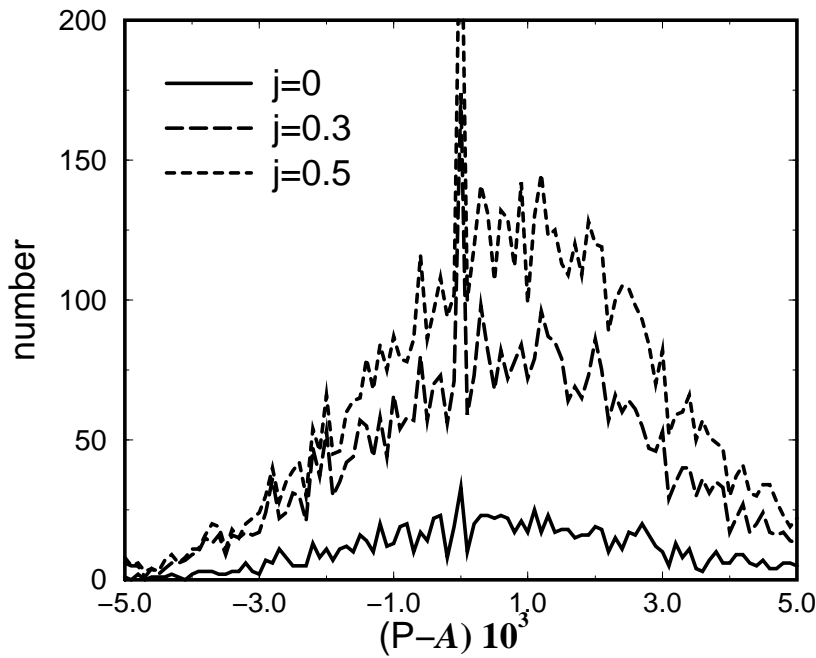} 
}
\vspace{-.8cm} 
\caption{ The distribution of the plaquette around its average value 
   ${\cal A}$ for several values of the noise suppression factor $j$.} 
\label{fig:2} 
\end{figure} 
Let us study the efficiency with which the action in (\ref{eq:10}) 
suppresses the statistical error of the average value of the plaquette. 
For this purpose we consider the probability distribution of finding 
a particular value of the plaquette $P_{\mu \nu }$ in the interval 
\be 
[ {\cal A} \, - \, 5 \times 10 ^{-3}, \; {\cal A} \, + \, 
5 \times 10 ^{-3} ] \; . 
\label{eq:13} 
\en 
From the numerical point of view, we proceed as follows: we 
divide the above interval in bins of width $10^{-4}$, and 
calculate the lattice average of the plaquette in a particular 
heat bath step. We then count the number of average values which 
correspond to a certain bin. 
We evaluated $1140$ heat bath steps. The numerical result is shown in 
figure \ref{fig:2} for $\beta =2.1$ and $j=0.5$. 
One clearly observes that the data points are 
strongly grouped around the corresponding average value ${\cal A}$ for 
large values of the noise suppression factor $j$. 
In addition, one observes a sharp peak at $P_{\mu \nu } = {\cal A}$. 
Whether this peak is an artifact due to corrections to the action of 
order $a^8$ or whether the peak is necessary to reproduce the 
standard Yang-Mills action (\ref{eq:11}) in the scaling limit 
$\beta \rightarrow \infty $, is not clear to us.

\subsection{The SU(2) mass gap } 

In this subsection, we will numerically estimate the SU(2) mass gap 
from the plaquette-plaquette correlation function in order to 
demonstrate how  the action (\ref{eq:10}) works in practice. 
The purpose of this subsection is two-fold: Firstly, we want 
to show that the mass gap obtained here is in agreement 
with the high precision measurements~\cite{tep87} which employ high 
statistics and an improvement of the signal--to--noise--ratio using the 
fuzzing technique. Secondly, we will compare the results at 
several values of $\beta $ in order to estimate the contribution of 
the ''irrelevant'' operators off the continuum limit. 
We are aware of the problem that the overlap of the plaquette with 
glue-ball wave-function is small~\cite{for86}. For high precision 
measurements, one should therefore employ non-local operators. 
Performing the noise suppression for the case of these operators, however, 
might be numerically costly. For these first investigations, we 
therefore confine ourselves to the study of correlations of the plaquette. 

Furthermore, we use the source method~\cite{fal85} to calculate the 
correlation function 
of the plaquettes. For this purpose, we estimate the average plaquette 
at the origin from 
\be 
W[\eta ] \; = \; \frac{ \int {\cal D} U_\mu \; \sum _{\mu \nu } 
P_{\mu \nu }(0) \; \exp \left\{ -S + \sum_{\{x\} \mu \nu } 
\eta _x P_{\mu \nu }(x) \right\} }{ 
\int {\cal D} U_\mu \; \; \exp \left\{ -S + \sum_{\{x\} \mu \nu } 
\eta _x P_{\mu \nu }(x) \right\} } 
\label{eq:14} 
\en 
where $S$ is the action (\ref{eq:10}). Let $P(x)$ denote 
$\sum _{\mu \nu } P_{\mu \nu }(x)$. It is straightforward to verify 
that the functional derivative of $W[\eta ]$ with respect 
to the source $\eta (x)$ yields the desired correlation function, i.e. 
\be 
C(t) \; = \; 
\frac{ \delta W[\eta ] }{ \delta \eta (x) } \vert _{\eta =0} \; = \; 
\langle P(x) P(0) \rangle \; - \; \langle P(x) \rangle \langle P(0) 
\rangle \; . 
\label{eq:15} 
\en 
In practice, we are interested in the correlation of the plaquette 
in time direction implying that one chooses $\eta (x) = \eta (t)$. 
In fact, one simulates two statistical ensembles. One ensemble is 
generated with the inverse temperature, the other is obtained by 
setting the inverse temperature of the time slice $t=0$ to 
$\beta + \eta $ leaving the remaining $\beta $-values unchanged~\cite{fal85}. 
In both ensembles, the average plaquette 
is obtained as a function of time. The correlation function (\ref{eq:15}) 
is obtained by approximating the functional derivative in (\ref{eq:15}) 
by the difference of the expectation values of the plaquette of the two  
ensembles. 

\begin{figure}[t]
\centerline{ 
\epsfxsize=12cm
\epsffile{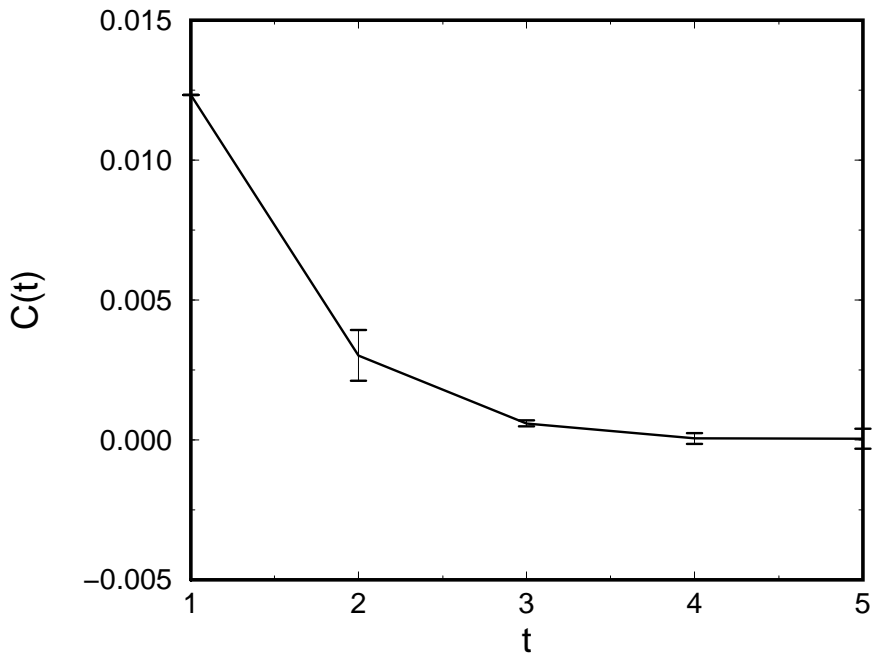} 
}
\vspace{-.8cm} 
\caption{ The correlation function (\protect{\ref{eq:15}}) as a function 
   of time $t$ in units of lattice spacings for $\beta =1.9$ and 
   $j=0.5$. } 
\label{fig:3} 
\end{figure} 
We use $1140$ heat bath steps to extract the average plaquette in both 
ensembles. A typical result for the correlation function as a function 
of time is shown in figure \ref{fig:3}, where the noise suppression factor 
is set to $j=0.5$. The inverse temperature is $\beta =2.1$ 
guaranteeing that the systems are in the scaling region (see 
left picture of figure \ref{fig:1}). One clearly observes an 
exponential decay of the correlation, where the slope of 
$\ln C(t) $ provides the SU(2) mass gap in units of the lattice spacing. 

It is interesting to compare the value of the mass gap, obtained with and 
without noise suppression, for $\beta $-values corresponding to the middle 
and the onset of the scaling window, respectively. In the latter case, 
the contribution of ''irrelevant operators'' to the mass gap should be 
more pronounced compared with the former case. 
The numerical results are summarized in the following table. 

\bigskip
\centerline{ 
\begin{tabular}{|c|c|c|c|c|c|} \hline 
$\beta $ & $j$ & $\kappa a^2 $ & $L$ & $m \, a$ & $ m / \sqrt{\kappa } $ \\ 
\hline 
$2.1$    & $0$ & $0.577$     & $3.5 \, $fm & $1.40 \pm 0.14$ & 
$1.85 \pm 0.2$ \\ 
$1.9$    & $0.5$ & $0.577$   & $3.5 \, $fm & $1.56 \pm 0.23$ & 
$2.06 \pm 0.3$ \\ 
$2.3$    & $0$ & $0.21$      & $2.1 \, $fm & $1.36 \pm 0.03 $ & 
$2.97 \pm 0.06 $ \\ 
$2.1$    & $0.5$ & $0.21$      & $2.1 \, $fm & $1.46 \pm 0.06 $ & 
$3.25 \pm 0.13 $ \\ 
$2.4$    & $0$ & $0.126$     & $1.6 \, $fm & $1.43 \pm 0.08$ & 
$4.04 \pm 0.3$ \\ 
$2.2$    & $0.5$ & $0.115$   & $1.5 \, $fm & $1.41 \pm 0.07$ & 
$4.14 \pm 0.2$ \\ \hline 
\end{tabular} 
} 
\bigskip

The error bars indicate the uncertainty due to statistical fluctuations. 
They are extracted from the fit of the correlation function $C(t)$ to 
the function $const. \, \exp \{ - mt \} $. 
The string tension $\kappa $ sets the scale. We use $\kappa = 440 \, $MeV. 
$L$ is the extension of our lattice in each direction. One observes 
that the discrepancy between the values $m/ \sqrt{\kappa }$ with and without 
noise suppression decreases, if the ensemble turns towards the 
continuum limit, i.e. $\kappa a^2$ decreases. This shows that 
the influence of the ''irrelevant'' operators, which distinguishes our 
action (\ref{eq:15}) from the standard Wilson action diminishes. 
Note, however, that at physical lattice sizes where the influence 
of the ''irrelevant'' operators is small, finite size effects might 
play a role. This would imply that a large number of lattice points 
(here we use $10^4$ lattice points) 
is necessary for high precision measurements of the SU(2) mass gap. 
The numerical result for the mass gap in the scaling region is in 
agreement with the results of~\cite{tep87}.

\section{Conclusions} 

We have shown that constraining the plaquette to its average value 
can be understood as composite field renormalization. The parameter $j$, 
which enters the action, acts as a renormalization constant 
absorbing the divergences arising from the composite nature of fields. 
These divergences are responsible for the small 
signal--to--noise--ratio 
in correlation functions of plaquettes. The action (\ref{eq:10}) 
with noise suppression differs from the standard Wilson action 
up to a shift in $\beta $ only by ''irrelevant operators'', i.e. 
both actions coincide in the naive continuum limit. 

We have numerically studied the new action (\ref{eq:10}) which implements 
the noise suppression on a coarse grained lattice consisting of 
$10^4$ lattice points. We have used a $\beta $ independent noise suppression 
factor $j$. Other choices are possible and perhaps more convenient 
depending on the type of correlation function which is under 
consideration. 

We have extracted the Creutz ratios from the numerical data with and without 
noise suppression and have established a scaling window in both cases. 
We have numerically confirmed that the action (\ref{eq:10}) 
reproduces the scaling of the standard Wilson action with 
$\beta $ shifted to a lower value in agreement with the analytical 
result. 

The SU(2) mass gap has been estimated from the plaquette--plaquette 
correlation function. We have found that the contributions from the 
''irrelevant'' operators to the screening mass decrease with 
increasing values of $\beta $. The goal of the noise suppression 
has mainly been the reduction of the statistical fluctuations of the 
''background'', on top of which the signal exists. 

For high precision measurements of glue-ball masses, one should use 
correlation functions of operators which have a larger overlap with 
the glue-ball wave function than the plaquettes. In addition, 
''perfect'' actions will help to extrapolate to the continuum limit. 
A generalization of the noise suppression introduced in the present paper 
to either case seems feasible. In the case of the ''perfect'' 
actions, one has to ensure that the noise suppression term does not 
spoil the correct ultra-violet behavior exploited by the ''perfect'' 
action technique.

\end{document}